\documentclass[fleqn,12pt,twoside]{article}
\usepackage[headings]{espcrc1}

\readRCS
$Id: espcrc1.tex,v 1.2 2004/02/24 11:22:11 spepping Exp $
\usepackage{graphicx}
\usepackage[figuresright]{rotating}

\newcommand{\AmS}{{\protect\the\textfont2
  A\kern-.1667em\lower.5ex\hbox{M}\kern-.125emS}}

\title{Nuclear Astrophysics in Rare Isotope Facilities}

\author{C.A. Bertulani\address[MCSD]{Department of Physics and Astronomy, Texas A\&M University, Commerce,
TX 75429, USA}%
        \thanks{This work was partially supported by the U.S. DOE
grants DE-FG02-08ER41533 and DE-FC02-07ER41457
(UNEDF, SciDAC-2), the Research Corporation under
Award No. 2009-7123.}}
       
\runtitle{Nuclear Astrophysics in Rare Isotope Facilities}
\runauthor{C.A. Bertulani}

\begin{document}

\maketitle

\begin{abstract}
Nuclear reactions in stars are difficult to measure directly in the
laboratory at the small astrophysical energies. In recent years
indirect methods with rare isotopes have been developed and applied to extract
low-energy astrophysical cross sections \cite{BG09}. 
\end{abstract}

{\bf A. Elastic scattering and ${\bf (p,\ p^{\prime })}$ reactions}

Elastic proton scattering has been one of the
major sources of information on the matter distribution of unstable
nuclei in radioactive beam facilities. The extended matter
distribution of light-halo nuclei ($^8$He, $^{11}$Li,  $^{11}$Be,
etc.) was clearly identified in elastic scattering experiments
\cite{Neu95,Kor97}. Information on the matter distribution of many
nuclei important for the nucleosynthesis in inhomogeneous Big Bang
and in r-process scenarios could also be obtained. 
In (p, p') scattering one obtains information on the excited states
of the nuclei 
\cite{alam}.

{\bf B. Transfer reactions}

Transfer reactions $A(a,b)B$ are effective when a momentum matching
exists between the transferred particle and the internal particles
in the nucleus. Thus, beam energies should be in the range of a few
10-100 MeV per nucleon \cite{angela}. Low energy reactions of
astrophysical interest can be extracted directly from breakup
reactions $A+a \longrightarrow b+c+B$ by means of the Trojan
Horse technique  \cite{Bau86}. If the
Fermi momentum of the particle $x$ inside $a=(b+x)$ compensates for
the initial projectile velocity $v_a$, the low energy reaction
$A+x=B+c$ is induced at very low (even vanishing) relative energy
between $A$ and $x$. Very successful results
using this technique have been reported 
\cite{Con07}.

The Asymptotic Normalization
Coefficient (ANC) technique relies on fact that the amplitude for
the radiative capture cross section $b+x\longrightarrow a+ \gamma$
is given by $M=\left<I_{bx}^a({\bf r_{bx}})|{\cal O}({\bf r_{bx}})|
\psi_i^{(+)}({\bf r_{bx}})\right>$, where $I_{bx}^a=\left<\phi_a(\xi_b, \ \xi_x,\ {\bf %
r_{bx}}) |\phi_x(\xi_x)\phi_b(\xi_b)\right>$ is the integration over the
internal coordinates $\xi_b$, and $\xi_x$, of $b$ and $x$,
respectively. For low energies, the overlap integral $I_{bx}^a$ is
dominated by contributions from large $r_{bx}$. Thus, what matters
for the calculation of the matrix element $M$ is the asymptotic
value of $I_{bx}^a\sim C_{bx}^a \ W_{-\eta_a, 1/2}(2\kappa_{bx}
r_{bx})/r_{bx}$, where $C_{bx}^a$ is the ANC and $W$ is the
Whittaker function. This
method  has been used with great success for many reactions of
astrophysical interest 
\cite{Muk90,Tri06},
with the advantage of avoiding the treatment of the screening
problem \cite{Con07}.

{\bf C. Coulomb Excitation and Dissociation}

Coulomb excitation in radioactive beam facilities has been
very successful to extract information on electromagnetic
properties of nuclear transitions of astrophysical interest
\cite{Glas01}. A reliable extraction of useful nuclear
properties from Coulomb excitation experiments at intermediate
and high energies requires a proper treatment of special relativity
\cite{Ber05}. 

The (differential, or angle integrated) Coulomb breakup cross
section for $a+A\longrightarrow b+c+A$  is directly proportional to the photo-nuclear cross section
$\sigma_{\gamma+a\ \rightarrow\ b+c}^{\pi\lambda}(\omega)$  for the multipolarity ${\pi\lambda}$ 
and photon energy $\omega$. Time reversal allows one to deduce the radiative
capture cross section $b+c\longrightarrow a+\gamma$ from $\sigma_{\gamma+a\ \rightarrow\ b+c}%
^{\pi\lambda}(\omega)$ \cite{BBR86}. The method has been used
successfully in a number of reactions of interest for astrophysics \cite{Tohru,EBS05}.

The contribution of the nuclear breakup has been examined by several
authors (see, e.g. \cite{BN93}). $^8$B has a small proton separation
energy ($\approx 140$ keV). For such loosely-bound systems it had
been shown that multiple-step, or higher-order effects, are
important \cite{BB93}.

{\bf D. Charge exchange reactions}

Charge exchange reactions induced in (p, n) reactions are often used
to obtain values of Gamow-Teller matrix elements, $B(GT)$, which
cannot be extracted from beta-decay experiments. This approach
relies on the similarity in spin-isospin space of charge-exchange
reactions and $\beta$-decay operators. As a result of this
similarity, the cross section $\sigma($p,\ n$)$ at small momentum
transfer $q$ is closely proportional to $B(GT)$ for strong
transitions:
$
{d\sigma\over dq}(q=0)=KN_D|J_{\sigma\tau}|^2 B(\alpha)
,
$
where $K$ is a kinematical factor, $N_D$ is a distortion factor (accounting for
initial and final state interactions), $J_{\sigma\tau}$ is the Fourier transform
of the effective nucleon-nucleon interaction, and $B(\alpha=F,GT)$ is the reduced transition
probability for non-spin-flip, $B(F)=
(2J_i+1)^{-1}| \langle f ||\sum_k  \tau_k^{(\pm)} || i \rangle |^2$,
and spin-flip,
$B(GT)=
(2J_i+1)^{-1}| \langle f ||\sum_k \sigma_k \tau_k^{(\pm)} || i \rangle |^2$, transitions.

The above relation, valid for one-step processes, was proven to
work rather well for (p,n) reactions (with a few exceptions). For
heavy ion reactions the formula might not work so well. This has
been investigated in refs. \cite{Len89,Ber93}. In ref. \cite{Len89}
it was shown that multistep processes involving the physical
exchange of a proton and a neutron can still play an important role
up to bombarding energies of 100 MeV/nucleon. Ref. \cite{Ber93} uses
the isospin terms of the effective interaction to show that
deviations from the above formula is common under many
circumstances. As shown in ref. \cite{Aus94}, for important GT
transitions whose strength are a small fraction of the sum rule the
direct relationship between $\sigma($p,\ n$)$ and $B(GT)$ values
also fails to exist. Similar discrepancies have been observed
\cite{Wat85} for reactions on some odd-A nuclei including $^{13}$C,
$^{15}$N, $^{35}$Cl, and $^{39}$K and for charge-exchange induced by
heavy ions \cite{St96}. 

But charge-exchange reactions such as (p,n), ($^{3}$He,t)
and heavy-ion reactions (A,A$\pm$1) provide information on the
$B(F)$ and $B(GT)$ values needed for astrophysical purposes. This is
one of the major research areas in radioactive beam facilities and
has been used successfully  \cite{chex09}.

{\bf E. Knock-out reactions}

Single-nucleon knockout reactions with heavy ions, at intermediate
energies and in inverse kinematics, have become a specific and
quantitative tool for studying single-particle occupancies and
correlation effects in the nuclear shell model \cite{Gregers,han03}. The experiments
observe reactions in which fast, mass $A$, projectiles collide
peripherally with a light nuclear target producing residues with
mass $(A-1)$ \cite{han03}. The final state of the target and that of
the struck nucleon are not observed, but instead the energy of the
final state of the residue can be identified by measuring
coincidences with decay gamma-rays emitted in flight.

New experimental approaches based on knockout reactions have been
developed and shown to reduce the uncertainties in astrophysical
rapid proton capture (rp) process calculations due to nuclear data.
This approach utilizes neutron removal from a radioactive ion beam
to populate the nuclear states of interest. In the first case
studied  \cite{clement}, $^{33}$Ar,
excited states were measured with uncertainties of several keV. The
2 orders of magnitude improvement in the uncertainty of the level
energies resulted in a 3 orders of magnitude improvement in the
uncertainty of the calculated $^{32}$Cl(p,$\gamma$)$^{33}$Ar rate
that is critical to the modeling of the rp process. 

{\bf F. Theoretical efforts}

 Recent works
\cite{VZ05,Mic04} are paving the way toward a microscopic
understanding of the many-body continuum. A basic theoretical
question is to what extent we know the form of the effective
interactions for threshold states. It is  hopeless that these
methods can be accurate in describing high-lying states in the
continuum. In particular, it is not worthwhile to pursue this
approach to describe direct nuclear reactions.

A less ambitious goal can be achieved in the coming years by using
the Resonating Group Method (RGM) or the Generator Coordinate Method
(GCM) \cite{QN09}. These is a set of coupled integro-differential equations of
the form \hfil\break
$
\sum_{\alpha'} \int d^3 r'
\left[
H^{AB}_{\alpha\alpha'}({\bf r,r'})-EN^{AB}_{\alpha\alpha'}({\bf r,r'})
\right]
g_{\alpha'}({\bf r'})=0,\label{RGM}
$\hfil\break
where $H^{AB}_{\alpha\alpha'}({\bf r,r'})=\langle \Psi_A(\alpha,{\bf
r})|H| \Psi_B(\alpha',{\bf r'}) \rangle$ and
$N^{AB}_{\alpha\alpha'}({\bf r,r'}) =\langle \Psi_A(\alpha,{\bf r})|
\Psi_B(\alpha',{\bf r'}) \rangle$. In these equations $H$ is the
Hamiltonian for the system of two nuclei (A and B) with the energy
$E$, $\Psi_{A,B}$ is the wavefunction of nucleus A (and B), and
$g_{\alpha}({\bf r})$ is a function to be found by numerical
solution of the above equation, which describes the relative motion of A
and B in channel $\alpha$. Full antisymmetrization between nucleons
of A and B is implicit. 

A simpler method was adopted in ref.
\cite{NBC05,NBC06}, where an excellent agreement was found with the momentum distributions in knockout
reactions of the type $^8$B$+A\longrightarrow \ ^7$Be$+X$ obtained
in experiments at MSU and GSI facilities. The astrophysical S-factor
for the reaction  $^{7}$Be$($p$,\gamma)^{8}$B was also calculated
and excellent agreement was found with the experimental data in both
direct and indirect measurements \cite{NBC05,NBC06}. 

Field theories adopt a completely independent approach for nuclear physics
calculations in which the concept of nuclear potentials is not used. The basic
method of field theories is to start with a Lagrangian for the fields. From this
Lagrangian one can ``read'' the Feynman diagrams and make practical
calculations, after taking care of well-known complications such as
regularization and renormalization. In nuclear astrophysics, this theory has been applied to $ np\rightarrow d\gamma$
for big-bang nucleosynthesis \cite{Chen99,Rup00}; $\nu d$ reactions for
supernovae physics \cite{KR99} and the solar $pp$ fusion process
\cite{But01}. EFT has also been used to deduce observables in reactions with
halo nuclei  and loosely bound states, with promising applications to
astrophysics \cite{BHK02,BHK03,HWK08}.

\end{document}